\documentclass[12pt,a4paper]{article}
\usepackage[T2A]{fontenc}
\usepackage[cp1251]{inputenc}

\begin{document}
\begin{center}
{\bfseries NOTES ON THE NATURE OF SO-CALLED INTRINSIC  SYMMETRIES.}
 \vskip 5mm {\bf
O.S.~Kosmachev} \vskip 5mm
{\it Joint Institute for Nuclear Research, 141980, Dubna, Russia} \\
{\it E-mail: kos@theor.jinr.ru}
\end{center}

\begin{minipage}{130mm}
\centerline{\bf Abstract} We show that in some cases intrinsic symmetries are
peculiar conversion or reflection of external symmetries. They form the
structure of the microparticles and then determine interactions of these
particles. \\
{\bf Key-words:} groups, irreducible representations, discrete symmetries,
spin, wave equations.
\end{minipage}
\begin{center}
\large\bf {Introduction}
\end{center}

The notion "intrinsic property" has the
dual sense. First of all it express some inalienable property, which is
inherent in object in any case. But sometimes it reflects a presence  of a
hidden property or insufficiently known one. Every manifestation of intrinsic
or any other property is result of interaction with external objects. Therefore
intrinsic properties are relative notions.

   In this article exhaustive analysis of some mathematical objects is
represented. All they have physical sense such as the electron spin, lepton
wave equations. Exhaustive analysis means so full consideration, that it left
no possibilities for a continuation of the mathematical analysis and therefore
it eliminates the presence  of some additional physical characteristics or
interpretations. Intrinsic symmetries in below considered examples are the
peculiar conversion or reflection of the space-time symmetries and discrete
ones.
\section{Pauli group and spin of the electron}
   The problem of the spin nature can not be considered solved in spite of a long
history of spin concept \cite{UG} and its successful mathematical formalization
\cite{P1} for the electron. At first \cite{UG} electron was called "spinning".
Then Pauli \cite{P1} named it "magnetic". A question on rotation of point like
particle is beyond reasonable understanding, therefore spin properties were
referred to internal or proper characteristics of the particle . Evidently, the
physical picture does not become more understandable.

   For a start let us appeal to the generally accepted assumption made by Pauli
that the spin is the proper angular momentum of the electron, having the
quantum nature and not related to motion of the particle as a whole. This
definition has not changed at present. In the case of the electron the proper
angular momentum is known to be described by Pauli $\sigma$-matrices:
\begin{displaymath}
\sigma_x= \left(\begin{array}{cc}
0 & 1 \\
1 & 0 \\
\end{array}\right);\quad
\sigma_y= \left(\begin{array}{cc}
 0 & -i\\
i & 0\\
\end{array}\right);\quad
\sigma_z= \left(\begin{array}{cc}
1  & 0 \\
0 & -1 \\
\end{array}\right).\quad
\end{displaymath}
It is easily checked that $\sigma$-matrices  generate the 16-th order group
just Pauli group \cite{BL}. We will denote this group as $d_{\gamma}$. The
group has ten conjugate classes. The center of the group contains four
elements. The group has eight one-dimensional and two non-equivalent
two-dimensional irreducible representations (IR's). The rank of the group is
equal to 3. This means that all three Pauli $\sigma-$matrices are necessary to
generate the group.

Let us introduce the following notations \cite{K8}:
\begin{displaymath}
\sigma_{z}\sigma_{y}\equiv a_1, \qquad \sigma_{x}\sigma_{z}\equiv a_2, \qquad
\sigma_{y}\sigma_{x}\equiv a_3
\end{displaymath}
and
\begin{displaymath}
\sigma_{x}\equiv b_{1}, \qquad \sigma_{y}\equiv b_{2}, \qquad \sigma_{z}\equiv
b_{3}
\end{displaymath}
It can be shown that
\begin{equation}
b_{1}=a_{1}c,\quad b_{2}=a_{2}c, \quad b_{3}=a_{3}c,
\end{equation}
where $c$ is  one of four $( I, -I, iI, -iI)$ elements of the group center and
 $c=\sigma_{x}\sigma_{y}\sigma_{z}=iI$. Here $I$ is the unit
$2\times 2$ matrix. This means that operators $a_1, a_2, a_3$ are connected
with operators $b_1, b_2, b_3$ by simple relations for the given irreducible
representation
\begin{equation}
b_{1}=ia_{1},\quad b_{2}=ia_{2}, \quad b_{3}=ia_{3}.
\end{equation}
It can also be noted that
\begin{equation}
a_2a_1a_2^{-1}=a_1^{-1}=a_1^3,\qquad a_1a_2\equiv a_3,\qquad a_1^2=a_2^2=a_3^2.
\end{equation}
This means that elements $a_1, a_2$ generate the quaternion subgroup \cite{K1}.
Let us denote it as - $Q_2[a_1,a_2]$.

Assuming that elements of the group $d_{\gamma}$ are generators of some
algebra, we obtain the following commutation relations (CR) for the elements of
the algebra (infinitesimal operators of the proper Lorentz group
representation)
\begin{equation}
\begin{array}{lll}
[a_1,a_2]=2a_3,    & [a_2,a_3]=2a_1,    & [a_3,a_1]=2a_2,  \\
\nonumber  [b_1,b_2]=-2a_3,   & [b_2,b_3]=-2a_1, & [b_3,b_1]=-2a_2,
\\  \nonumber  [a_1,b_1]=0,        & [a_2,b_2]=0,
&[a_3,b_3]=0,
\\  \nonumber  [a_1,b_2]=2b_3,     & [a_1,b_3]=-2b_2,   &   \\
\nonumber  [a_2,b_3]=2b_1,    & [a_2,b_1]=-2b_3,   &    \\
\nonumber  [a_3,b_1]=2b_2,    & [a_3,b_2]=-2b_1.   &
\end{array}
\end{equation}

The obtained commutative relations coincide with commutative relations of the
infinitesimal matrices of the proper homogeneous Lorentz group \cite{N} to the
factor 2 common for all equalities. Due to construction of commutative relation
(4), all six operators $a_1, a_2, a_3$ and $b_1, b_2, b_3$ have a definite
physical meaning.

It follows from the first row of commutative relations (CR's) (4) that elements
$a_1, a_2, a_3$ and all their products form the subgroup of 3-dimensional
rotations. As it follows from the derivation of commutative relations \cite{N},
$ b_{1}=\sigma_{x}, \quad b_{2}=\sigma_{y}, \quad b_{3}=\sigma_{z}$ have the
sense of infinitesimal operators of Lorentz transformations.

Taking into account the anticommutation of the operators $b_1, b_2, b_3$, the
second upper row of commutative relations (4) takes the form:
\begin{equation}
b_1b_2=-a_3,\quad   b_2b_3=-a_1,\quad   b_3b_1=-a_2,
\end{equation}
All three equalities express in infinitesimal form the rotation by some fixed
angel of one inertial system with respect to another at their relativistic
motion \cite {M}. Obviously, upon deviation from uniform rectilinear motion,
this effect has a more complex nature. Upon transition to regular repeated
motion, for example,  to orbital motion, the rotation also becomes regular,
i.e., is manifested as rotation . That is why only the sum of orbital momentum
and the spin can be the integral of motion of the particle moving along orbit,
rather than orbital momentum and the spin separately.

Thus the analysis of $\sigma$-matrix group on the base of CR's (4) which are
the direct corollary of the Lorentz transformations demonstrates that so-called
proper momentum of the particle with the spin equal 1/2 is the consequence of a
definite character of motion of this particle, which is not free or
rectilinear. This conclusion is in agreement with the well known fact. It is
impossible to measure magnetic moment of the electron related with spin
momentum, if it moves freely \cite{MM}.

The explicit form of the operators $a_1, a_2, a_3$ and $b_1, b_2, b_3$ for
irreducible representations allows to evaluate two weight numbers $(l_0, l_1)$,
which specify uniquely irreducible representations of the Lorentz group.
Calculation of the eigenvalue for the standard $\sigma$-matrices yields
$l_0=1/2$. We see, that first weight number ($l_0$) coincides with spin value.

The value of the first weight number is determined formally by operators $
a_1,a_2,a_3$, i.e. by the subgroup of three-dimensional rotations. But
generation of the spin rotation is impossible without relativity, as it follows
from the above mentioned. It is undoubtedly truly to correlate spin with quite
a definite quantum number, if the quantum numbers are interpreted as indices of
groups \cite{WG}. But it is an unprovable assumption to endow the spin notion
with a physical value which exists separately from the motion of the electron
as a whole.

Thus two sides of the spin conception has been demonstrated. The first is
related to the form of equation. It determines the fermion or boson type of
particles. As it is shown above this side stems from  the three-dimensional
rotation subgroup. This yields the strict fixing of the spin value as the
integer or half-integer constant, rather than physical quantity. It is this
side of spin notion  that is present in the spin definition proposed by Pauli.

The second side is related to the occurrence of the physical quantity of the
spin momentum of the electron (and corresponding magnetic momentum) in the
interaction resulting in nonuniform motion. If the motion becomes periodical,
repeated, we obtain the particle spin as the physical quantity. In this
manifestation (according to the second row of commutation relations (4)) spin
is no more a strictly fixed constant. The circumstance that the first side
related to the form of equation is initial obligatory and independent on the
second becomes fundamental. The second side is realized only in the presence of
the nonuniform motion and depends on the first one in its manifestations and
details. This can influence significantly the analysis of compound systems or
particles with internal structure.

From the generally accepted formalism of the spin equal 1/2 without  any
additional assumption, we obtain one of the irreducible representation of the
Lorentz group and, as a corollary, a physical interpretation of Pauli
$\sigma$-matrices. Strictly speaking, it is applicable for description of
electrons or objects, whose structures are not taken into account.

Consequent and more detailed examination of Pauli group structure
($d_{\gamma}$) show, that it has duality. It means that apart from subgroup
$Q_2[a_1,a_2]$ it contains one more subgroup of  eight order - $q_2[a_1,a'_2]$.
Defining relations between the generators are the same for both groups.
Difference is the order of generators. Both generators of $Q_2[a_1,a_2]$ has
fourth order. One generator($a_1$) of $q_2[a_1,a'_2]$ has  fourth order and
another is of second one. Commutation relation for $q_2[a_1,a'_2]$ (Lie
algebra) has the form:
\begin{equation}
[a_1,a'_2]=2a'_3,\qquad     [a'_2,a'_3]=-2a_1,\qquad    [a'_3,a_1]=2a'_2,
\end{equation}
where $a_1=\sigma_{z}\sigma_{y}, \quad a'_2=a_{2}c, \quad a_3=a_1a'_2,\quad
c=\sigma_{x}\sigma_{y}\sigma_{z}=iI$

 Let us call $q_2[a_1,a'_2]$ as a quaternion group of the second kind. It
is not difficult show, that $Q_2[a_1,a_2]$ is related to $SU(2)$ with
 $detU=1$ whereas $q_2[a_1,a'_2]$ is related to $SU(2)$ with $detU=-1$.

 If we extend $q_2[a_1,a'_2]$ by the same element as previously
 $c=\sigma_{x}\sigma_{y}\sigma_{z}=iI$, we obtain following commutation
 relations:
 \begin{equation}
\begin{array}{lll}
[a_1,a'_2]=2a'_3,    & [a'_2,a'_3]=-2a_1,    & [a'_3,a_1]=2a'_2,  \\
\nonumber [b'_1,b'_2]=-2a'_3,   & [b'_2,b'_3]=2a_1, & [b'_3,b'_1]=-2a'_2,
\\  \nonumber  [a_1,b'_1]=0,        & [a'_2,b'_2]=0,
& [a'_3,b'_3]=0,
\\  \nonumber  [a_1,b'_2]=2b'_3,     & [a_1,b'_3]=-2b'_2,   &   \\
\nonumber [a'_2,b'_3]=-2b'_1,    & [a'_2,b'_1]=-2b'_3,   &    \\
\nonumber  [a'_3,b'_1]=2b'_2,    & [a'_3,b'_2]=2b'_1,   &
\end{array}
\end{equation}
where $b'_1=a_1c, \quad b'_2=a'_2c,\quad b'_3=a'_3c$

These relations are differed from those written above (4). We will connect them
with group $f_{\gamma}$, taken into account that $f_{\gamma}$ and $d_{\gamma}$
are isomorphic.

The representation (7) is called (P)-conjugate with respect to $d_{\gamma}$,
since distinctions appear at the level of the 3-dimensional rotation subgroup,
i.e. at the first row. The transition from (4) to (7)
 is equivalent to the following change
$a_2\rightarrow ia'_2$. Due to the definition of $a_3$ we obtain
$a_3\rightarrow ia'_3$. All further deviation from commutation relations (4) in
more lower rows are the consequences of this primary change. In this case the
quaternion subgroup $Q_2[a_1,a_2]$ transforms into $q_2[a_1,a'_2]$.

As a result, three spatial directions are not equivalent for $q_2[a_1,a'_2]$,
or the so-called asymmetry between the left and right is observed. The first
weight number is equal to $l_0=1/2$ for $q_2[a_1,a'_2]$ only for  $a_1$. The
number $l_0$ is obtained pure imaginary for other two operators ($a'_2, a'_3$),
i.e. they have no physical meaning for the three-dimensional rotation subgroup.

Similar non-equivalence is observed also for $b'_1, b'_2, b'_3$ which is an
equivalent of the Pauli $\sigma$-matrix P-conjugate representation. They
explicit form for $f_\gamma$ group is
\begin{displaymath}
b'_1= \left(\begin{array}{cc}
1 & 0 \\
0 & -1 \\
\end{array}\right);\quad
b'_2= \left(\begin{array}{cc}
 0 & 1\\
-1 & 0\\
\end{array}\right);\quad
b'_3= \left(\begin{array}{cc}
0 & -i \\
-i & 0 \\
\end{array}\right).\quad
\end{displaymath}
It lead provided some conditions to the spin orientation along the particle
momentum. Evidently, it is impossibly to add something more to analysis of
Pauli group.
\section{Lepton wave equations}

   Next examples, confirming previous one, are the different kinds of lepton wave
equations. All they are formulated in the frame work unique approach on the
base of unified mathematical formalism.

{\bf Algorithm of the stable lepton equations} was obtained by means of
exhaustive analysis of Dirac equation \cite{K7}. All completeness of
information on Dirac equation is determined by matrices $\gamma_{\mu}
(\mu=1,2,3,4.)$. They generate group of 32 order \cite{L} later
$D_{\gamma}(II)$. Determining relations for this group have the form \cite{D}
\begin{equation}
\gamma_{\mu}\gamma_{\nu} + \gamma_{\nu}\gamma_{\mu}=2\delta_{\mu \nu}, \quad
(\mu, \nu=1,2,3,4).
\end{equation}
It was established that $D_{\gamma}(II)$ contains two nonisomorphic subgroup of
16 order only. The first is above examined subgroup $d_{\gamma}$ and the second
was denoted by $b_{\gamma}$. Lie algebra has the form for $b_{\gamma}$:
\begin{equation}
\begin{array}{lll}
[a_1,a_2]=2a_3,       &[a_2,a_3]=2a_1,       &[a_3,a_1]=2a_2,  \\
\nonumber [b''_1,b''_2]=2a_3,     &[b''_2,b''_3]=2a_1, &[b''_3,b''_1]=2a_2,  \\
\nonumber [a_1,b''_1]=0, &[a_2,b''_2]=0,         &[a_3,b''_3]=0,  \\  \nonumber
[a_1,b''_2]=2b''_3      &[a_1,b''_3]=-2b''_2,    &   \\   \nonumber
[a_2,b''_3]=2b''_1,     &[a_2,b''_1]=-2b''_3,    &    \\  \nonumber
[a_3,b''_1]=2b''_2,     &[a_3,b''_2]=-2b''_1.    &
\end{array}
\end{equation}
Transition from (4) to (9) is realized by change
\begin{equation}
b_k\rightarrow b''_k=ib_k \quad (k=1,2,3).
\end {equation}
Subgroup $b_{\gamma}$ in contrast to $f_{\gamma}$ is T-conjugate connected
component with respect to $d_{\gamma}$.

Then was obtained fourth connected component, i.e. (TP)-conjugate connected
component with respect to $d_{\gamma}$. This  group was designet $c_{\gamma}$.
The corresponding Lie algebra has the form:
\begin{equation}
\begin{array}{lll}
[a_1,a'_2]=2a'_3,       &[a'_2,a'_3]=-2a_1, &[a'_3,a_1]=2a'_2,  \\
\nonumber [b^*_1,b^*_2]=2a'_3, &[b^*_2,b^*_3]=-2a_1, &[b^*_3,b^*_1]=2a'_2,  \\
\nonumber [a_1,b^*_1]=0, &[a'_2,b^*_2]=0,         &[a'_3,b^*_3]=0,
\\  \nonumber [a_1,b^*_2]=2b^*_3,      &[a_1,b^*_3]=-2b^*_2,    &
\\   \nonumber [a'_2,b^*_3]=-2b^*_1,     &[a'_2,b^*_1]=-2b^*_3,
&    \\  \nonumber [a'_3,b^*_1]=2b^*_2,     &[a'_3,b^*_2]=2b^*_1. &
\end{array}
\end{equation}
These four constituents are complete set for description  of any lepton
 wave equation.

 Algorithm of every concrete lepton equation is based on the following
 general requirements:
\begin{enumerate}
 \item The equations must be invariant and covariant under proper Lorentz
 transformations taken into account all four connected components.
 \item The equations must be formulated on the base of irreducible
 representations of the groups determining every lepton equation.
 \item Conservation of four-vector of probability current must be fulfilled and
 fourth component of the current must be positively defined.
 \item The lepton spin is supposed equal to 1/2.
 \item Every lepton equation must be reduced to Klein-Gordon equation.
 \end{enumerate}

    The basis for each equation is the appropriate group of $\gamma$-matrices.
    Each group of $\gamma$-matrices is produced by four generators. Three of
    them must anticommute, and the fourth may anticommute or commute with the
    first three ones. These requirements comprise necessary and sufficient
    conditions for the formulation of wave equation for a free stable lepton.
    None of stated requirements can be excluded without breaking the equation,
    the Dirac equation in particular. The latter served as a prototype  making
    it possible to formulate the algorithm. As a result, it has been found
    that each equation for stable lepton has its own unique structure.

    \begin{enumerate} \item Dirac equation --- $D_{\gamma}(II)$:
$d_{\gamma}, b_{\gamma}, f_{\gamma}$. \\ $In[D_{\gamma}(II)]=-1$   \item
Equation for doublet of massive neutrinos
нейтрино --- $D_{\gamma}(I)$: $d_{\gamma}, c_{\gamma}, f_{\gamma}$. \\
$In[D_{\gamma}(I)]=1$ \item Equation for quartet of massless neutrinos  ---
$D_{\gamma}(III)$: $d_{\gamma}, b_{\gamma}, c_{\gamma}, f_{\gamma}$. \\
$In[D_{\gamma}(III)]=0$ \item Equation for massless $T$-singlet ---
$D_{\gamma}(IV)$: $b_{\gamma}$.
\\ $In[D_{\gamma}(IV)]=-1$      \item Equation for massless $P$-singlet
--- $D_{\gamma}(V)$: $c_{\gamma}$\\ $In[D_{\gamma}(V)]=1$.
\end{enumerate}
   Here $d_{\gamma}, b_{\gamma}, f_{\gamma}, c_{\gamma}$ are subgroups of the
appropriate group of $\gamma$-matrices on which one of the four connection
components of the Lorentz group is realized. It is seen that the structure of
the equation enable to differentiate one equation from another. All of the
equations have no substructures allowing for a physical interpretation.For this
reason they are stable as the electrong. What is more, the proposed method
allows an assertion that there are no other stable leptons in the context of
the assumption mentioned above. It turn out possible due to theorem on three
types of irreducible  matrix groups \cite{L}. New kind of wave equation
invariant was used which is numerical characteristic taking one of three
possible values $\pm 1,0$. This value is denoted here as $In[D_{\gamma}]$. For
example, it is equal to $In[D_{\gamma}(II)]=-1$  for Dirac equation \cite{K3}.

{\bf Unstable leptons and their structures.} The problem of the lepton quantum
numbers which make them distinguishable is a cardinal problem for the lepton
sector. The first necessary step for it solving is formulation free state for
every lepton \cite{K10}. It has turned out that it is possible to obtain extra
lepton equation within the framework  of the previous assumption.

   The given task is solved by introducing  an additional (the fifth) generator
to produce a group of $\gamma$-matrices. It was found that there are three and
only three possible extensions. Each of them is equivalent to the introducing
of it own additional quantum numbers. In this case in new groups, there arise
substructures allowing for a physical interpretation in terms of stable
leptons. Therein lies their main distinction from the previous ones, making
them unstable.

The extension of the Dirac $\gamma$-matrix group $(D_{\gamma}(II))$ with the
help of one anticommuting generator $\Gamma_5$, such that $\Gamma_5^2=I$, leads
to a group $\Delta_1$ with a structural invariant $In[\Delta_1]=-1$. The
extension of the same group with the help of a generator, such that
$\Gamma_5'^2=-I$, yields a group $\Delta_3$ with a structural invariant
$In[\Delta_3]=0$. Finally, the extension of the group of $\gamma$-matrices of
the doublet neutrino $(D_{\gamma}(I))$ with the help of $\Gamma_5''^2=-I$
results in a group $\Delta_2$ with an invariant $In[\Delta_1]=1$.

The order all of the three groups is equal to 64, and each of the groups have
32 one-dimension irreducible representations and two nonequivalent
four-dimension ones. Besides, each is composed of three and only three
subgroups of the 32nd order, which are isomorphic to one of the five indicated
cases. But the composition of the subgroups of the 32bd order is unique in each
case.

{\bf Group $\Delta_1$} has following defined relations
\begin{equation}
\Gamma_{\mu}\Gamma_{\nu} + \Gamma_{\nu}\Gamma_{\mu}=2\delta_{\mu \nu}, \quad
(\mu, \nu=1,2,3,4,5)
\end{equation}
It follows from these relations that
\begin{equation}
 \Gamma_{6}\Gamma_{\mu}=\Gamma_{\mu}\Gamma_{6}, \quad \Gamma_{6}^2=I \quad
(\mu =1,2,3,4,5),
\end{equation}
where $\Gamma_{6} \equiv \Gamma_{1}\Gamma_{2}\Gamma_{3}\Gamma_{4}\Gamma_{5}$.

It means that $\Gamma_{6}$ is the center of the group. By this is means that,
as we go to an irreducible matrix representations, we can write $\Gamma_{6}=\pm
I$. When $\mu$ and $\nu$ run through the value 1,2,3,4 we obtain the Dirac
group. It can be shown starting from (12)that, in addition to the Dirac
subgroup, the $\delta_1$ contains two and only two subgroups of the 32nd order.
As result we arrive at the following structural decomposition:
\begin{equation}
\Delta_1 \{D_{\gamma}(II),\quad D_{\gamma}(III),\quad D_{\gamma}(IV)\}
\end{equation}
The relation (14), together with the structure invariant $In[\Delta_1]=-1$,
identifies the group $\Delta_1$, i.e., make its physical contents different
from the rest.

{\bf The group $\Delta_3$} results from the extension of the Dirac group by
means of similar defining relations. The only distinction is in the order of
the fifth generator $\Gamma_5$.
\begin{equation}
\begin{array}{ll}
\Gamma_{s}\Gamma_{t}+\Gamma_{t}\Gamma_{s}=2\delta_{st},  & (s,t=1,2,3,4),\nonumber\\
\Gamma_{s}\Gamma_{5}+\Gamma_{5}\Gamma_{s}=0,  & (s=1,2,3,4), \nonumber\\
\Gamma_{5}^2=-1.
\end{array}
\end{equation}
Hence it follows that
\begin{equation}
 \Gamma_{6} \equiv \Gamma_{1}\Gamma_{2}\Gamma_{3}\Gamma_{4}\Gamma_{5},\quad
 \Gamma_{6}\Gamma_{\mu}=\Gamma_{\mu}\Gamma_{6},\quad
(\mu =1,2,3,4,5).
\end{equation}
As before, $\Gamma_{6}$ is a group center $\Gamma_{6}^2=-I$. In this case the
matrix realization of representation leads to $\Gamma_6=\pm iI$.

The decomposition of the group has changed as follows:
\begin{equation}
\Delta_3 \{D_{\gamma}(II),\quad D_{\gamma}(I),\quad D_{\gamma}(III)\},
\end{equation}
which corresponds to the structural invariant $In[\Delta_3]=0$. The Dirac
subgroup is presented as before, but the composition as a whole has changed.

{\bf The group $\Delta_2$} is determined by the relations:
\begin{equation}
\begin{array}{ll}
\Gamma_{s}\Gamma_{t}+\Gamma_{t}\Gamma_{s}=2\delta_{st},  & (s,t=1,2,3),\nonumber\\
\Gamma_{s}\Gamma_{4}+\Gamma_{4}\Gamma_{s}=0,  & (s=1,2,3), \nonumber\\
\Gamma_{4}^2=-1.\nonumber \\
\Gamma_{u}\Gamma_{5}+\Gamma_{5}\Gamma_{u}=0,  & (u=1,2,3,4), \nonumber\\
\Gamma_{5}^2=-I.
\end{array}
\end{equation}
It follows that
\begin{equation}
 \Gamma_{6} \equiv \Gamma_{1}\Gamma_{2}\Gamma_{3}\Gamma_{4}\Gamma_{5},\quad
 \Gamma_{6}\Gamma_{\mu}=\Gamma_{\mu}\Gamma_{6}\quad
(\mu =1,2,3,4,5),
\end{equation}
 $\Gamma_{6}$ is an element of the group center and $\Gamma_{6}^2=I$.
 In the matrix realization, it takes the form $\Gamma_{6}=\pm I$.

The group decomposition differs from the two previous ones:
\begin{equation}
\Delta_2 \{D_{\gamma}(I),\quad D_{\gamma}(III),\quad D_{\gamma}(V)\},
\end{equation}
as does structure invariant $In[\Delta_2]=1$. Comparison of the $\Delta_2$
composition with the five above-cited equations for stable leptons shows that
the expression (20) contains only neutrino components.

The presence of the Dirac subgroup in the groups $\Delta_1$ and $\Delta_3$
implies that the latter are evident candidates for the description of unstable
charged leptons.

It should be note, that irreducible representation of $\Delta_2$-group was
obtained in pure real form. This is in agreement with above mentioned theorem
on irreducible matrix groups \cite{L}. The theorem states in particular that,
if  $In[D_{\gamma}]=1$, there exist a nonsingular matrix that transforms all of
matrices of the representation into real ones, which is what has been done.
Taken into account the content of $\Delta_2$-group, one can say, that it is the
basis for description of unstable massive neutrino. This situation is entirely
repeats the stable massive lepton one.
\begin{center}
\large\bf {Conclusion}
\end{center}
The method has a peculiar "stability" with respect to variations of the
generator orders. Their orders equal to four or two are restricted by
requirement to have the spin equal to 1/2. The extension of any of the five
groups to describe stable lepton by using the anticommuting generator result in
the three groups $\Delta_1, \Delta_2, \Delta_3$ and no others. It would be very
difficult to understand the structures of the unstable leptons without whole
knowledge of the stable leptons.

In all examined cases the simplest constituents are the connection components
of the Lorentz group in different combinations and nothing more. This
differences are the basis for manifestation various properties, quantum numbers
and so on due to interactions.

All examined cases are particular examples. But the method of the analysis
completely general, i.e. exhaustive structural analysis. It is create
prerequisites to believe that such examples will be multiply.


\begin{thebibliography}{99}
\bibitem{UG} G.E. Uhlenbeck, S. Goudsmit, "Ersetzung der
Hypothese vom unmechanischen Zwang durch eine Forderung bez\"uglich des inneren
Verhalters jedes einzelnen Elektrons", Naturwissenschaften, {\bf 13}, S.953
(1925). "Spinning electrons and the structure of spectra" Nature. {\bf 117},
p.264 (1926).
\bibitem{P1} W.Pauli, "Zur Quantenmechanik des magnetischen
Electrons", Zs f. Phys. {\bf B.43}. S.601 (1927).
\bibitem{BL} L.C.Biedenharn, J.D.Louck, {\it Angular momentum in quantum physics}
v.1 (Cambridge University Press, Cambridg, 1985).
\bibitem{K8} O.S.Kosmachev,
 "Spin as a Kinematical Effect of the Relativity",
Preprint OIYaI, P2-2005-6 (Dubna, 2005) [in Russian].
\bibitem{K1} O.S.Kosmachev,
 "A Physical Interpretation of Some Group Algebras", Letters Part. and Nucl.
{\bf v.1, N5}, p.50 (2004).
\bibitem{N} M.A.Na\v{i}mark, {\it Linear Representations of the
Lorentz Group} (Fizmatgiz, Moskow, 1958), p.88 [in Russian].
\bibitem{M} C. M\o ller {\it The Theory of the Relativity}
(Clarendon Press Oxford, 1972; Atomizdat, Moscow 1975, p.39).
\bibitem{MM} N.F.Mott, H.S.W.Massey, {\it The theory of Atomic Collisions}
(Clarendon Press, Oxford 1965; Mir, Moscow, 1969, p.204).
\bibitem{WG} G.Weyl, {\it The Theory of Groups and Quantum
Mechanics},(Dover Publications, 1931; Nauka, Moscow, 1986, p.16).
\bibitem{K7} O.S.Kosmachev, "Representations of Lorentz Group and Classification of
Stable Leptons," Preprint OIYaI R2-2006-6 (Dubna, 2006).[in Russian]
\bibitem{L} Lomont J.S. {\it Applications of finite groups} (Academic Press,
New York, London, 1959) P.40, P.51.
\bibitem{D} P.Dirac,
"The quantum theory of the electron" Proc.Roy. Soc. A {\bf vol.117},(1928)
P.610-624.
\bibitem{K3} O.S.Kosmachev,
 "On Invariants of Dirac Type Equations",
Preprint OIYaI, P2-2002-217 (Dubna, 2002).[in Russian]
\bibitem{K10} O.S.Kosmachev, "The Problem of Quantum Numbers of the Lepton
Sector", Preprint OIYaI P2-2008-73 (Dubna, 2008).[in Russian]

\end{thebibliography}
\end{document}